\documentclass[12pt]{article}
\usepackage{graphicx}
\usepackage[cp1251]{inputenc}

\begin{document}
 \noindent {\footnotesize\it   Astronomy Letters, 2020, Vol. 46, No 7, pp. 439--448.}

 \noindent
 \begin{tabular}{llllllllllllllllllllllllllllllllllllllllllllll}
 & & & & & & & & & & & & & & & & & & & & & & & & & & & & & & & & & & & & & \\\hline\hline
 \end{tabular}

  \vskip 0.5cm
  \centerline{\bf\large Study of the Galactic Rotation Based on Masers and Radio Stars}
  \centerline{\bf\large with VLBI Measurements of Their Parallaxes}
   \bigskip

   \bigskip
   \centerline
    {V. V. Bobylev$^{[1]}$\footnote{e-mail: vbobylev@gaoran.ru}
    O. I. Krisanova$^{[2]}$ and A. T. Bajkova$^{[1]}$ }

  \bigskip
  \centerline{\small\it $^{[1]}$ Pulkovo Astronomical Observatory, Russian Academy of Sciences,}

  \centerline{\small\it Pulkovskoe sh. 65, St. Petersburg, 196140 Russia}

  \centerline{\small\it $^{[2]}$
  St. Petersburg State University, Universitetskaya nab. 7/9, St. Petersburg, 199034 Russia}

 \bigskip
 \bigskip
 \bigskip

 {
{\bf Abstract}---Based on published data, we have produced a sample of 256 radio sources whose trigonometric
parallaxes and proper motions were measured by VLBI. This sample contains Galactic masers associated
with massive protostars and stars in active star-forming regions. It also includes young low-mass stars
from the Gould Belt region whose radio observations were performed in continuum. Based on this, most
complete sample of sources to date, we have estimated the velocities $U_\odot,V_\odot,W_\odot$ and the parameters of the angular velocity of Galactic rotation $\Omega_0,\Omega^{(1)}_0,\ldots,\Omega^{(4)}_0$ and obtained a new estimate of the distance from the Sun to the Galactic center, $R_0=8.15_{-0.20}^{+0.04}$~kpc. The parameters of the Galactic spiral density wave have been found from the series of radial, $V_R,$ and residual tangential, $\Delta V_{circ}$, velocities of stars. The amplitudes of the radial and tangential velocity perturbations are $f_R=7.0\pm0.9$~km s$^{-1}$ and $f_\theta=3.8\pm1.1$~km s$^{-1}$, the perturbation wavelengths are $\lambda_R=2.3\pm0.2$~kpc and $\lambda_\theta=2.0\pm0.4$~kpc, and the Sun’s phases in the spiral density wave are $(\chi_\odot)_R=-163^\circ\pm9^\circ$ and $(\chi_\odot)_\theta=-137^\circ\pm10^\circ$ for the adopted four-armed spiral pattern.
  }


 \subsection*{INTRODUCTION}
Using data on young objects allows important information about the kinematic properties of the
Galactic disk to be obtained. These objects include, for example, neutral hydrogen clouds at the tangential
points, whose line-of-sight velocities play an important role in constructing the rotation curve of
the Galaxy in its inner region. Classical Cepheids realizing an independent distance scale based on the
period–luminosity relation are important. Open star clusters and OB associations are also of interest.

Present-day astrometric VLBI measurements have allowed a very high accuracy of determining
the kinematic characteristics of maser sources to be achieved. For example, the error in the trigonometric
parallaxes of masers at a frequency of 22 GHz is, on average, $\sim$0.01 mas (milliarcseconds) and
$\sim$0.01 mas yr$^{-1}$ (milliarcseconds per year) for their proper motions, with the period of observations being about two years or more (Reid and Honma 2014a). At present, these ground-based VLBI measurements
of trigonometric parallaxes are more accurate than the Gaia satellite measurements (Prusti et al. 2016;
Brown et al. 2018).

The rotation parameters of the Galaxy and the parameters of its spiral structure were determined
using data on masers by Reid et al. (2014b, 2016, 2019), Bobylev and Bajkova (2013, 2014a), Rastorguev
et al. (2017), Honma et al. (2018), and Hirota et al. (2020). The greatest number of maser
sources have been measured in the Local Arm and, therefore, its parameters have been determined quite
reliably (Xu et al. 2013; Bobylev and Bajkova 2014c). The parameters of the Perseus spiral arm are also
determined reasonably well (Sakai et al. 2015; Reid et al. 2019).

Recently, the distance from the Sun to the Galactic center $R_0$, close to 8 kpc, has been determined
quite confidently using a large number of masers (Reid et al. 2019; Hirota et al. 2020). The linear circular
rotation velocity of the Sun around the Galactic center determined from masers is close to 240 km s$^{-1}$
(Rastorguev et al. 2017; Reid et al. 2019; Hirota et al. 2020). Such a velocity is typical of the youngest
Galactic disk objects. In Reid et al. (2019) the data on masers served to refine the orientation parameters of
the Galactic plane. On the whole, high expectations in refining the structural and dynamical parameters of
the Galaxy are associated with maser sources. For example, Honma et al. (2015) showed that $R_0$ and $V_0$
would be determined with errors $\sim$1\% using a sample of 500 masers.

The local rotation parameters of the Galaxy are already known quite well. This has been achieved
by using mass measurements of the trigonometric parallaxes and proper motions of stars from such catalogues
as Hipparcos (1997) and Gaia. Only maser sources currently allow the structure and kinematics
of the Galactic disk to be traced in a very wide range of Galactocentric distances $R.$ Classical Cepheids
can ``compete'' with masers, but these are already older stars and, for example, they are not that clearly
associated with the spiral structure.

The goals of this paper are (i) to create a database of radio sources whose trigonometric parallaxes and
proper motions have been measured by VLBI based on published data and (ii) to estimate the Galactic
rotation parameters using these data.

 \subsection*{METHOD}
 \subsubsection*{Basic Equations}
The following quantities are known from observations: the right ascension and declination $\alpha$ and $\delta$, the parallax $\pi$, the proper motions in right ascension and declination $\mu_{\alpha} \cos \delta$ and $\mu_{\delta}$, and the line-of-sight velocity $V_r.$ It is easy to pass from $\alpha$ and $\delta$ to the Galactic longitude and latitude $l$ and $b$; the parallax gives the heliocentric distance $r,$ because
$r=1/\pi$; these proper motions can be converted to the proper motions in the Galactic coordinate system,
$\mu_l\cos b$ and $\mu_b$. Thus, we know three stellar velocity components: $V_r$ and two tangential velocity components, $V_l=kr\mu_l \cos b$ and $V_b=kr\mu_b$, where $k=4.74$ and $V_r, V_l, V_b$ are expressed in km s$^{-1}$ (the proper motions are given in mas yr$^{-1}$ and the heliocentric distances are in kpc).

Consider a kinematic model of the Galaxy by assuming that the centroids move in circular orbits
around the symmetry axis of the Galaxy in planes parallel to its principal plane (i.e., the rotation velocity
does not depend on the height of an object $z$ above the disk plane). In 1924--1925 Bottlinger derived the formulas describing the influence of the circular rotation of the centroids on the observed line-of-sight, $V_r,$ and tangential, $\Delta V_{\tau}$, velocities of stars:
\begin{equation}\label{VrVt}
 \begin{array}{lll}
      \Delta V_r=R_0 (\Omega-\Omega_0)\sin l\cos b \\
 \Delta V_{\tau}=R_0(\Omega-\Omega_0)\cos l-\Omega r\cos b,
 \end{array}
\end{equation}
where $R_0$ is the distance from the Sun to the Galactocentric center, $\Omega (R)$ is the angular velocity of Galactic rotation, and $\Omega_0=\Omega (R_0)$ is the angular velocity of Galactic rotation at the solar circle. The function $\Omega (R)$ can be expanded into a Taylor series in powers of $(R-R_0):$
\begin{equation}\label{Omega}
 \begin{array}{lll}
 \Omega(R)=\Omega(R_0)+\Omega'(R_0)(R-R_0)+\Omega''(R_0)(R-R_0)^2/2!+\ldots.
 \end{array}
\end{equation}
Restricting ourselves to the $n$th derivative in the above relation and taking into account the fact that
the peculiar solar motion enters into the observed line-of-sight and tangential velocities, we can obtain
the following system of equations:
\begin{equation}\label{eqVr}
 \begin{array}{lll}
 V_r=-U_{\odot}\cos b\cos l-V_{\odot}\cos b\sin l -W_{\odot}\sin b\\
  + R_0(R-R_0)   \sin l \cos b\Omega_0 '
  + R_0(R-R_0)^2 \sin l \cos b\Omega_0 '' /2+\ldots\\
  + R_0(R-R_0)^n \sin l \cos b\Omega_0^{(n)} /n!,
 \end{array}
\end{equation}
\begin{equation}\label{eqVl}
 \begin{array}{lll}
 V_l= U_{\odot}\sin l-V_{\odot}\cos l-r\Omega_0 \cos b \\
 + (R-R_0)   (R_0\cos l-r\cos b)\Omega_0 '
 + (R-R_0)^2 (R_0\cos l-r\cos b)\Omega_0 ''/2 + \ldots \\
 + (R-R_0)^n (R_0\cos l-r\cos b)\Omega_0^{(n)}/n!,
 \end{array}
\end{equation}
\begin{equation}\label{eqVb}
 \begin{array}{lll}
V_b = U_{\odot} \cos l \sin b +V_{\odot} \sin l \sin b-W_{\odot} \cos b\\
  -R_0(R-R_0)   \sin l \sin b\Omega_0 '
  -R_0(R-R_0)^2 \sin l \sin b\Omega_0 ''/2 - \ldots\\
  -R_0 (R-R_0)^n\sin l \sin b\Omega_0^{(n)}/n!,
 \end{array}
\end{equation}
where $U_{\odot}, \, V_{\odot}, \, W_{\odot}$ are the peculiar velocity components of the sample of stars being analyzed relative to the local standard of rest (LSR), where the LSR is the point moving in a circular orbit around the Galactic center, directed along the axes of the rectangular Galactic coordinate system, and $\Omega_0^{(i)}$ is the $i$th derivative of the angular velocity at the solar circle.

Occasionally, for simplicity, the velocities $U_{\odot}, \, V_{\odot}, \, W_{\odot}$ are called the peculiar solar velocity components. However, the velocity component $V_{\odot}$ is known to be affected by such an effect as the asymmetric drift (centroid lagging), which is revealed when studying the motions of stars of various ages. In addition, the influence of the Galactic spiral density wave manifests itself in the velocities $U_{\odot}$ and $V_{\odot}$ when analyzing the youngest stars (Bobylev and Bajkova 2014b).

Equations (3)--(5) are solved simultaneously or individually for the parameters $(U_{\odot}, \, V_{\odot}, \, W_{\odot}, \, \Omega_0, \, \Omega_0 ', \, \ldots, \, \Omega_0^{(n)})$  at fixed expansion order $n$ and parameter $R_0.$ The angular velocity at the solar circle and its derivatives in combination with $R_0$ determine the pattern of the rotation curve in the solar neighborhood, because
\begin{equation}\label{Vcirc}
 \begin{array}{lll}
  V_{circ}=R\cdot\Omega (R)= R[\Omega_0+(R-R_0)\Omega_0'+(R-R_0)^2\Omega_0''/2!+\ldots].
 \end{array}
\end{equation}
It is also necessary to introduce the circular rotation velocity at the solar radius $V_0$ and two Oort constants, $A$ and $B:$
\begin{equation}\label{AB}
 \begin{array}{lll}
 V_0=     \Omega_0 R_0,\\
  A = -0.5\Omega_0' R_0,\\
  B =    -\Omega_0+A.
\end{array}
\end{equation}
The rectangular components of the stellar space velocities are calculated from the formulas
\begin{equation}\label{UVW}
 \begin{array}{lll}
U=V_r \cos l \cos b-V_l \sin l-V_b \cos l \sin b,\\
V=V_r \sin l \cos b+V_l \cos l-V_b \sin l \sin b,\\
W=V_r \sin b+V_b \cos b.
\end{array}
\end{equation}
Using $U$ and $V,$ the circular velocity $V_{circ}$ directed along the Galactic rotation is expressed as
\begin{equation}\label{UVW}
 V_{circ}=U\sin\theta +(V_0+V)\cos\theta,
\end{equation}
where the angle $\theta$ satisfies the relation $\tan\theta=-Y/X,$ where $X$ and $Y$ are the Galactocentric
rectangular coordinates of a star.

 \subsubsection*{Least-Squares Method}
To determine the parameters of the rotation curve, Eqs. (3)--(5) are solved simultaneously or individually
by the weighted least-squares method (LSM) with weights of the form
\begin{equation}\label{weights}
 \begin{array}{lll}
 w_r =S_0/\sqrt{S_0^2+\sigma_{V_r}^2}, \\
 w_l =S_0/\sqrt{S_0^2+\sigma_{V_l}^2}, \\
 w_b =S_0/\sqrt{S_0^2+\sigma_{V_b}^2},
 \end{array}
\end{equation}
where $\sigma_{V_r}, \, \sigma_{V_l}, \, \sigma_{V_b}$ are the dispersions of the corresponding
observed velocities and S0 is the ``cosmic'' dispersion. $S_0$ is comparable to the root-mean-square
residual $\sigma_0$ (the error per unit weight) that is calculated when solving the conditional equations
(3)--(5) and is taken in this paper to be 12 km s$^{-1}$.

However, in this approach it is necessary to fix the nonlinear parameter $R_0.$ By now there are a
number of studies in which the mean distance from the Sun to the galactic center is deduced based on its
individual determinations in the last decade by independent methods. For example, $R_0=8.0\pm0.2$~kpc
(Vall\'ee 2017a), $R_0=8.3\pm0.2$\,(stat.)$\pm0.4$\,(syst.)~kpc (de Grijs and Bono 2017), or $R_0=8.0\pm0.3$~ kpc $(2\sigma)$ (Camarillo et al. 2018).

The highly accurate measurements of $R_0$ made in recent years are also worth noting. For example,
Abuter et al. (2019) found $R_0=8.178\pm0.013$ (stat.)$\pm0.022$\,(syst.) kpc by analyzing a 16-year-long series of observations of the motion of the star S2 around the supermassive black hole $Sgr$A$^*$ at the Galactic center. Based on an independent analysis of the orbit of the star S2, Do et al. (2019) found $R_0=7.946\pm0.050$\,(stat.)$\pm0.032$\,(syst.)~kpc. Taking into account the listed studies, in this paper
we take the distance from the Sun to the Galactic center to be $R_0=8.0\pm0.15$~kpc when solving the
conditional equations by the weighted linear LSM. To eliminate the rough error, we solve the equations
at least in two iterations with the elimination of large residuals according to the $3\sigma$ criterion.

Consider a system of conditional equations in general form:
\begin{equation}
y = f(X, \, \textbf{a}),
\end{equation}
where $X$ are the known quantities, y are the measurements with weights $w,$ a is the vector of sought-for
values, and $f$ is the specified function. Let $M$ be the number of sought-for parameters and $N$ be
the number of equations in the system and $N>M.$ Then, the dimension of the vectors $y$ and $w$ will be $N.$

One of the methods for solving such a system consists in a numerical search for the minimum of the
sum of the squares of the residuals:
\begin{equation}
 \sigma^2 (\textbf{a})=
 \frac{1}{N_{free}}\sum_{j=1}^N w_j\,(y_j-f(x_j,\textbf{a}))^2,
\end{equation}
where $N_{free}=N-M.$

For our system of equations (3)--(5) the vector of sought-for parameters is represented as
\begin{equation}
\textbf{a} = (R_0,  U_0, V_0, W_0, \Omega_0, \Omega_0', \ldots, \Omega_0^{(n)}).
\end{equation}
$M=n+5,$ where $n$ is the order of the expansion in terms of $\Omega_0$. $N$ will be equal to the number of
objects multiplied by the number of components of the observed velocities involved in the system being
solved (for example, when deriving the parameters only from the components $V_l$ and $V_b,$ Eqs. (4) and (5),
the total number of equations in the system will be $2\,*$  number of objects), $f$ is described by one of Eqs. (3)--(5), and the weights $w_j$ are in the form (9).

In our case, the only nonlinear parameter is $R_0 \equiv a_1$. Therefore, we may exclude this parameter by
fixing it. Let us introduce the notation: $(a_1)_0 \equiv \min \, \sigma^2 (a_1), \; \sigma_0^2=\sigma^2 ((a_1)_0)$. When fixing $a_1,$ the system becomes linear and we can apply the linear LSM. The minimum of $(a_1)_0$ is found from the dependence $\sigma^2 (a_1)$ and the LSM solution with it will give us a point estimate of the remaining parameters: $(a_2)_0, \, \ldots, \, (a_M)_0$.

The errors (confidence intervals) of the parameters can be determined using the statistic
\begin{equation}
\chi (\textbf{a}) = \sum_{j=1}^N \left( \frac{y_j - f_j (\textbf{a})}{\sigma_j} \right)^2,
\end{equation}
where $\sigma_j$ are the true errors of the measurements $y_j,$ which are assumed to be known. Let us introduce the notation:
\begin{equation}
\chi_0^2 \equiv \min \, \chi^2 (\textbf{a})=\chi^2 (\textbf{a}_0),
\end{equation}
\begin{equation}
 \textbf{a}_0 = [(a_1)_0, \, \ldots, \, (a_M)_0],
\end{equation}
\begin{equation}
\chi_1^2 (a_m) \equiv \min_{a_m=const} \, \chi^2 (\textbf{a}).
\end{equation}
At the $1\sigma$ significance level the boundaries of the confidence interval are determined from the equation
\begin{equation} \label{chi_eq}
\chi_1^2 (a_m) = \chi_0^2 + 1.
\end{equation}
Then, the final estimates of the parameters can be represented as $a_m={(a_m)_0}_{-\sigma_m^{-}}^{+\sigma_m^{+}}$, where $\sigma_m^{-}, \, \sigma_m^{+}$ are
the roots of (17).

In reality, $\sigma_j$ are unknown, while the weights $w_j$ are known. Their scale can be corrected using $\sigma_0,$ the mean error per unit weight:
\begin{equation}\label{sigmaj}
w_j = \tilde{\sigma}_j^{-2}; \quad \sigma_j = \sigma_0 \tilde{\sigma}_j= \frac{\sigma_0}{w_j},
\end{equation}
where $\sigma_0^2 \equiv \min \sigma^2 (\textbf{a})=\sigma^2 (\textbf{a}_0)$. With $\sigma_j$ calculated from Eq. (18), Eq. (17) can be rewritten as
\begin{equation}
\sigma_1^2 (a_m) = \sigma_0^2 \left( 1 + \frac{1}{N_{free}} \right),
 \end{equation}
\begin{equation}\label{sigma_general}
 \sigma_1^2 (a_m) = \min_{a_m=const} \sigma^2 (\textbf{a}).
\end{equation}
Equation (20) is solved for all parameters $a_m$.

\begin{figure}[t]
{\begin{center}
  \includegraphics[width=0.5\textwidth]{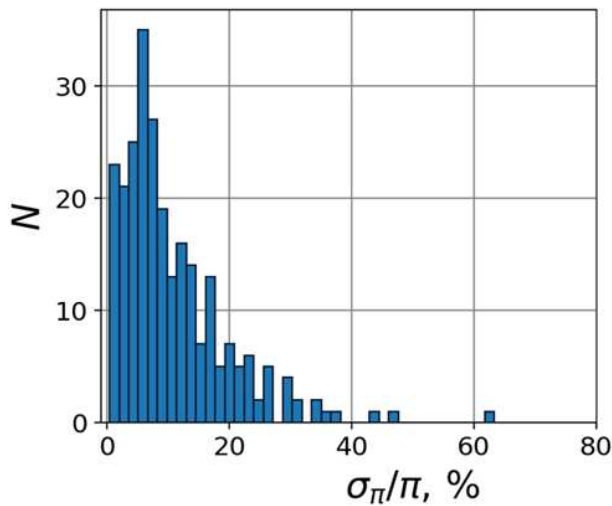}
 \caption{
 The distribution of relative parallax errors in the original sample of masers.
  } \label{masers_parallaxes}
\end{center}}
\end{figure}
\begin{figure}[t]
{\begin{center}
  \includegraphics[width=0.6\textwidth]{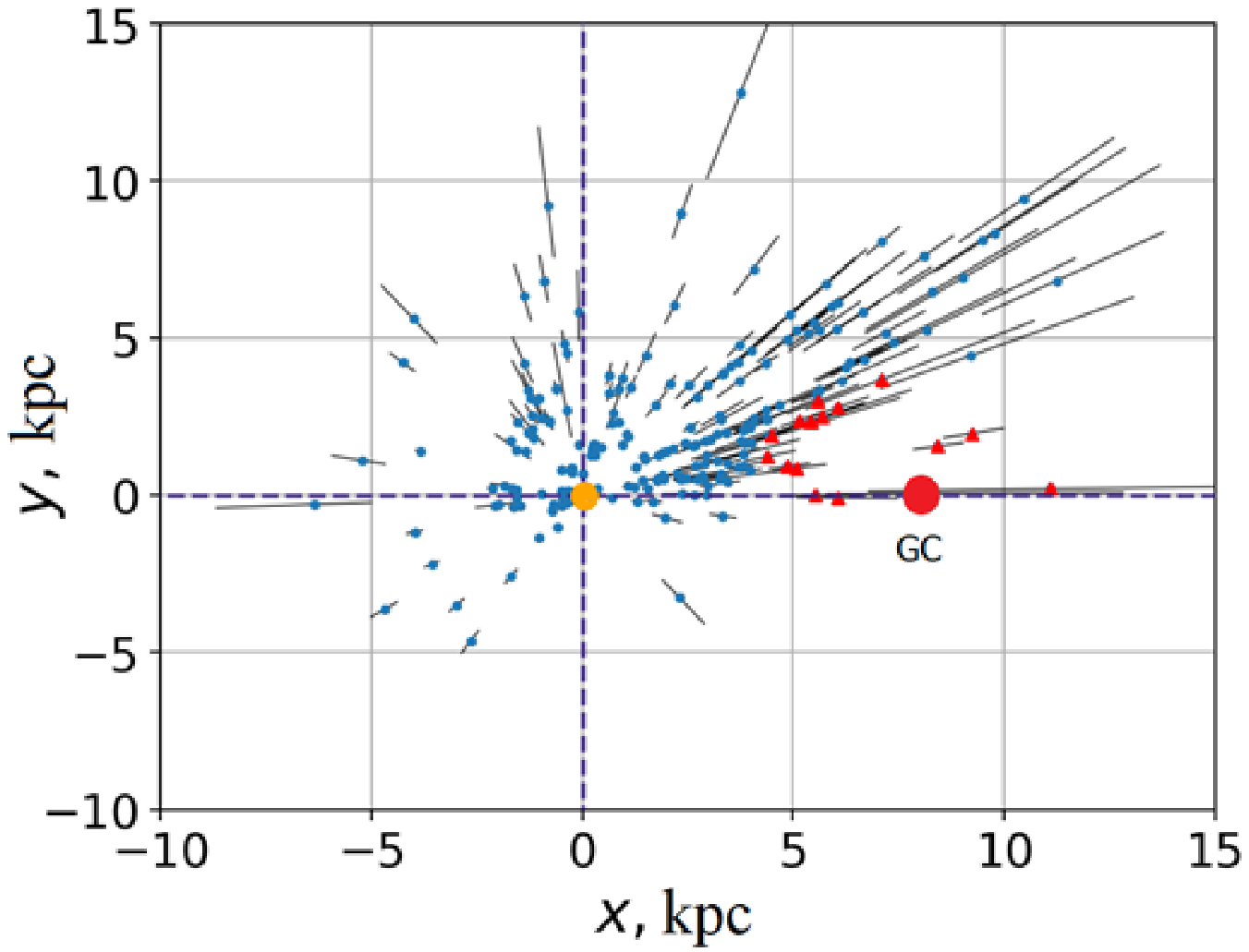}
 \caption{
The distribution of masers in projection onto the $xy$ plane. The orange point corresponds to the Sun’s position, while the red circle corresponds to the Galactic center (GC).
  } \label{masers_xyz}
\end{center}}
\end{figure}

 \subsubsection*{Spectral Analysis}
The influence of the spiral density wave in the radial, $V_R,$ and residual tangential, $\Delta V_{circ}$, velocities is periodic with an amplitude $\sim$10 km s$^{-1}$. According to the linear density wave theory (Lin and Shu 1964), it is described by the following relations:
 \begin{equation}
 \begin{array}{lll}
       V_R =-f_R \cos \chi,\\
 \Delta V_{circ}= f_\theta \sin\chi,
 \label{DelVRot}
 \end{array}
 \end{equation}
where
 \begin{equation}
 \chi=m[\cot(i)\ln(R/R_0)-\theta]+\chi_\odot
 \end{equation}
is the phase of the spiral density wave ($m$ is the number of spiral arms, $i$ is the pitch angle of the spiral
pattern, $\chi_\odot$ is the radial phase of the Sun in the spiral density wave); $f_R$ and $f_\theta$ are the amplitudes of the radial and tangential velocity perturbations, which are assumed to be positive.

We apply a spectral (periodogram) analysis to study the periodicities in the velocities $V_R$ and $\Delta V_{circ}$. The wavelength $\lambda$ (the distance between adjacent spiral arm segments measured along the radial direction) is calculated from the relation
\begin{equation}
 2\pi R_0/\lambda=m\cot(i).
 \label{a-04}
\end{equation}
Let there be a series of measured velocities $V_{R_n}$ (these can be the radial, $V_R,$ or tangential, $\Delta V_{circ}$, velocities), $n=1,\dots,N$, where $N$ is the number of objects. The objective of our spectral analysis is to extract a periodicity from the data series in accordance with the adopted model describing a spiral density wave with parameters $f,$ $\lambda$~(or $i)$ and $\chi_\odot$.

Having taken into account the logarithmic behavior of the spiral density wave and the position angles
of the objects $\theta_n$, our spectral analysis of the series of velocity perturbations is reduced to calculating the square of the amplitude (power spectrum) of the standard Fourier transform (Bajkova and Bobylev 2012):
\begin{equation}
 \bar{V}_{\lambda_k} = \frac{1} {N}\sum_{n=1}^{N} V^{'}_n(R^{'}_n)
 \exp\biggl(-j\frac {2\pi R^{'}_n}{\lambda_k}\biggr),
 \label{29}
\end{equation}
where $\bar{V}_{\lambda_k}$ is the $k$th harmonic of the Fourier transform with wavelength $\lambda_k=D/k$, $D$ is the period of the series being analyzed,
 \begin{equation}
 \begin{array}{lll}
 R^{'}_{n}=R_0\ln(R_n/R_0),\\
 V^{'}_n(R^{'}_n)=V_n(R^{'}_n)\times\exp(jm\theta_n).
 \label{21}
 \end{array}
\end{equation}
The sought-for wavelength $\lambda$ corresponds to the peak value of the power spectrum Speak. The pitch angle of the spiral density wave is found from Eq. (23). We determine the perturbation amplitude and phase by
fitting the harmonic with the wavelength found to the observational data.

As a result, our approach consists of two steps: (i) the construction of a smooth Galactic rotation
curve and (ii) a spectral analysis of the radial, $V_R,$ and residual tangential, $\Delta V_{circ}$, velocities.

 \subsection*{DATA}
The maser sources are stars with vast envelopes in which the pumping effect arises. These can be both young stars and protostars of various masses and old stars, for example, Miras. In this paper we selected only young objects closely associated with active star-forming regions.

Our list of data on maser sources with measured trigonometric parallaxes combines two large compilations,
those by Reid et al. (2019) and Hirota (2020), including several results published by various authors
in 2020. Reid et al. (2019) collected information about 199 masers that were observed at various frequencies
within the BeSSeL (Bar and Spiral Structure Legacy Survey\footnote{http://bessel.vlbi-astrometry.org}) project. Hirota et al. (2020) presented a catalogue of 99 sources that were observed at 22\,GHz within the Japanese VERA (VLBI Exploration of Radio Astrometry\footnote{http://veraserver.mtk.nao.ac.jp}) program. Note that there is a high percentage of common measurements between the samples by Reid et al. (2019) and Hirota et al. (2020). We also added several new parallax determinations for a number of maser sources in the outer spiral arm (Sakai et al. 2019) and the refined parameters of the
source V838 Mon (Ortiz-Le\'on et al. 2020).

Apart from the maser sources, highly accurate VLBI measurements of the trigonometric parallaxes
and proper motions were performed for a number of low-mass (T Tauri) stars in continuum in the Gould
Belt region. These observations were carried out within the GOBELINS (Gould's BeltDistances Survey)
program. We used the data on about 30 such stars from Kounkel et al. (2017), Galli et al. (2018),
and Ortiz-Le\'on et al. (2018). All these stars are no farther than 500 pc from the Sun.

Our final sample contains 256 radio sources with known positions and space velocities. Figure 1
presents the distribution (histogram) of relative parallax errors in this sample. To derive the kinematic
parameters, we excluded all of the masers with a Galactoaxial distance $R$ less than 4 kpc, where there
is a tangible influence of the Galactic bar. The final sample obtained after that contains 239 objects.

The distribution of masers in projection onto the Galactic $xy$ plane is presented in Fig. 2. The blue dots
mark the masers remaining in the final sample and the red triangles mark those excluded from consideration.
The orange dot corresponds to the Sun's position and the red circle corresponds to the Galactic center. The
maser error distances are also shown on the graph. Since the main observations have been performed so
far only from the Earth's northern hemisphere, only half of the Galactic plane is filled.

\begin{table}[t]
\caption{The kinematic parameters found}
\begin{center}
\begin{tabular}{|l|r|r|r|r|}
\hline
Parameters & $n=1$ & $n=2$ & $n=3$ & $n=4$ \\
\hline
$R_0,$ kpc & $9.38_{-0.14}^{+0.27}$ & $8.15_{-0.09}^{+0.10}$ & $8.15_{-0.20}^{+0.04}$
           & $8.15_{-0.20}^{+0.04}$  \\
$U_{\odot},$ km s$^{-1}$ & $4.33_{-1.55}^{+1.53}$ & $7.64_{-1.25}^{+1.24}$ & $7.79_{-1.27}^{+1.23}$
                  & $7.79_{-1.26}^{+1.23}$ \\
$V_{\odot},$ km s$^{-1}$ & $6.99_{-1.58}^{+1.54}$ & $13.64_{-1.26}^{+1.23}$ &
                    $15.04_{-1.25}^{+1.24}$ & $15.09_{-1.27}^{+1.22}$ \\
$W_{\odot},$ km s$^{-1}$  & $8.57_{-1.49}^{+1.48}$ & $8.57_{-1.23}^{+1.18}$
                   & $8.57_{-1.23}^{+1.18}$ & $8.57_{-1.23}^{+1.18}$ \\
$\Omega_0,$ km s$^{-1}$ kpc$^{-1}$  & $28.94_{-0.42}^{+0.41}$ & $29.16_{-0.34}^{+0.33}$ & $29.01_{-0.34}^{+0.33}$ & $29.00_{-0.34}^{+0.33}$ \\
$\Omega_0^{(1)},$ km s$^{-1}$ kpc$^{-2}$  & $-2.804_{-0.075}^{+0.075}$ & $-4.086_{-0.069}^{+0.068}$ & $-3.901_{-0.069}^{+0.068}$ & $-3.927_{-0.069}^{+0.068}$ \\
$\Omega_0^{(2)},$ km s$^{-1}$ kpc$^{-3}$ & --- & $0.717_{-0.032}^{+0.032}$ & $0.831_{-0.032}^{+0.032}$ & $0.848_{-0.032}^{+0.031}$ \\
$\Omega_0^{(3)},$ km s$^{-1}$ kpc$^{-4}$ & --- & --- & $-0.104_{-0.019}^{+0.018}$ & $-0.084_{-0.019}^{+0.018}$ \\
$\Omega_0^{(4)},$ km s$^{-1}$ kpc$^{-5}$ & --- & --- & --- & $-0.017_{-0.014}^{+0.013}$ \\
$\sigma_0,$ km s$^{-1}$ & $15.84$ & $12.87$ & $12.82$ & $12.83$ \\
\hline
\end{tabular}
\label{t2-masers}
\end{center}
\end{table}

 \subsection*{RESULTS AND DISCUSSION}
 \subsubsection*{Galactic Rotation Parameters}
Figure 3 presents the profiles of the function $\sigma^2 (R_0)$ for the $\Omega_0$ expansion orders under consideration: $n = {1, 2, 3, 4}.$ 

Table 1 gives the estimates of the sought-for parameters with their confidence intervals. Figure 4 presents the rotation curves for three cases of the $\Omega_0$ expansion: $n = {2, 3, 4}.$ As can be seen from Fig. 4,
an increase in the unknowns being determined leads to an expansion of the confidence region, which is
especially noticeable at large $R.$ Out of the three cases presented in the figure, it is better to choose the
case where the rotation curve is closest to the flat one. Thus, the curve in Fig. 4a goes upward too early (at
$R\sim13$ kpc). The curve in Fig. 4c differs too much from the flat one at large $R.$ For example, the case
with three $(n=3)$ derivatives of the angular velocity is better suited for obtaining the residual velocities
of Cepheids $\Delta V_{circ}$ with the goal of their spectral analysis (Fig. 4b). As can be seen from the table, the error per unit weight for this case has a minimum value, $\sigma_0=12.82$ km s$^{-1}$.

For $n=3$ we can estimate the following quantities by assuming the errors to be symmetric: 
 $V_0=236.4\pm4.4$ km s$^{-1}$ for $R_0 = 8.15\pm0.12$ kpc, 
 $\Omega_\odot=30.51\pm0.34$ km s$^{-1}$ kpc$^{-1}$, where $\Omega_\odot=\Omega_0+V_\odot/R,$
and $V_\odot=12.24$ km s$^{-1}$ is taken from Sch\"onrich et al. (2010). Here, $\Omega_\odot$ is the angular velocity of the Sun around the Galactic center.

\begin{figure}[t]
{\begin{center}
  \includegraphics[width=0.5\textwidth]{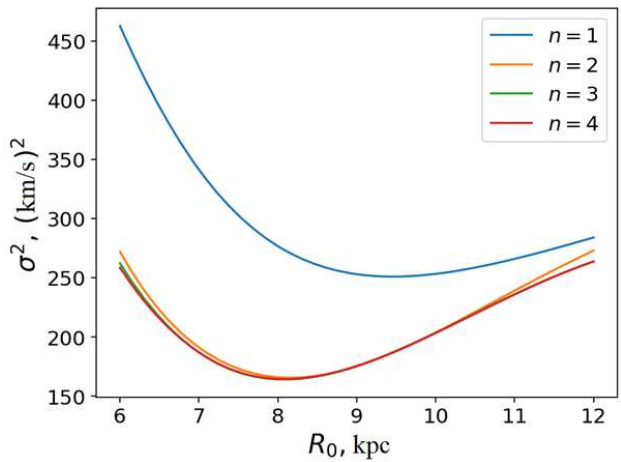}
 \caption{
 The $\sigma^2$ profile as a function of $R_0$ for various orders of the expansion in terms of $\Omega_0.$
  } \label{R0-masers}
\end{center}}
\end{figure}

Based on 130 Galactic masers with measured trigonometric parallaxes, Rastorguev et al. (2017)
found the solar velocity components $(U_\odot,V_\odot)=(11.40,17.23)\pm(1.33,1.09)$ km s$^{-1}$ and the following
parameters of the Galactic rotation curve:
 $\Omega_0=28.93\pm0.53$ km s$^{-1}$ kpc$^{-1}$, 
 $\Omega^{'}_0=-3.96\pm0.07$ km s$^{-1}$ kpc$^{-2}$, 
 $\Omega^{''}_0=0.87\pm0.03$ km s$^{-1}$ kpc$^{-3}$,
and $V_0=243\pm10$ km s$^{-1}$ for the value of $R_0=8.40\pm0.12$ kpc found.

Based on a sample of 147 masers, Reid et al. (2019) found the following values of the two most important
kinematic parameters: $R_0=8.15\pm0.15$ kpc and 
 $\Omega_\odot=30.32\pm0.27$ km s$^{-1}$ kpc$^{-1}$, where $\Omega_\odot=\Omega_0+V_\odot/R.$ 
 The velocity $V_\odot=12.24$ km s$^{-1}$ was taken from Sch\"onrich et al. (2010). These authors used the
expansion of the linear Galactic rotation velocity into a series.

Based on a similar approach, Hirota et al. (2020) obtained the following estimates by analyzing
99 masers that were observed within the VERA program:
 $R_0=7.92\pm0.16$~(stat.)~$\pm0.3$~(syst.) kpc and
  $\Omega_\odot=30.17\pm0.27$~(stat.)~$\pm$0.3~(syst.) km s$^{-1}$ kpc$^{-1}$,
where $\Omega_\odot=\Omega_0+V_\odot/R,$ and the velocity $V_\odot=12.24$ km s$^{-1}$ was also taken from Sch\"onrich et al. (2010).

It is also interesting to note the paper by Ablimit et al. (2020), where $\sim$3500 classical Cepheids with
proper motions from the Gaia DR2 catalogue were used to construct the Galactic rotation curve. The
Galactic rotation curve was constructed from the Cepheids of this sample in the range of distances $R:$
4--19 kpc. The circular rotation velocity of the solar neighborhood was found with a very high accuracy to
be $V_0 = 232.5\pm0.9$ km s$^{-1}$.

\begin{figure}[t]
{\begin{center}
  \includegraphics[width=0.66\textwidth]{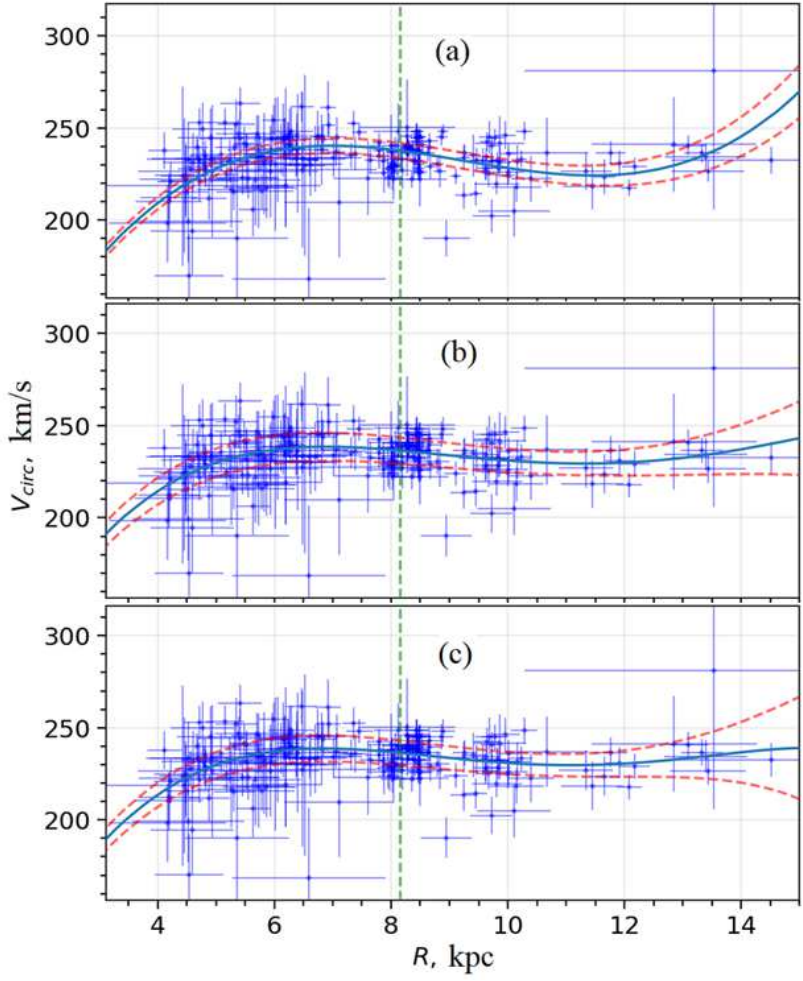}
 \caption{
 Maser rotation velocity $V_{circ}$ versus $R.$ The rotation curves are presented for the angular velocity expansion orders $n=2$ (a), 3 (b), and 4 (c). The $1\sigma$ confidence regions are indicated. The vertical line marks the Sun's position.}
\end{center}}
\end{figure}

 \subsubsection*{Spiral Density Wave Parameters}
For our spectral analysis we used 239 stars in the range of distances $R$ from 4 to 15 kpc. For our
further analysis of the residual tangential velocities we chose the rotation curve with three derivatives ($n=3$ from Table 1). Based on the deviation from it, we calculated the residual circular velocities $\Delta V_{circ}$. All our calculations here were carried out with $R_0=8.15$ kpc.

Based on the series of radial, $V_R,$ and residual tangential, $\Delta V_{circ}$, velocities, we found the following parameters for the adopted four-armed spiral pattern $(m = 4):$ the amplitudes of the radial and tangential velocity perturbations are $f_R=7.0\pm0.9$ km s$^{-1}$ and $f_\theta=3.8\pm1.1$ km s$^{-1}$, the perturbation wavelengths are $\lambda_R=2.3\pm0.2$ kpc and $\lambda_\theta=2.0\pm0.4$ kpc, and the phases of the Sun in the spiral density wave are $(\chi_\odot)_R=-163^\circ\pm9^\circ$ and $(\chi_\odot)_\theta=-137^\circ\pm10^\circ$.

For the four-armed spiral pattern for $\lambda_R=2.3\pm0.2$ kpc we can calculate the pitch angle from Eq. (23),
$i=10.2\pm1.0^\circ$. This value is consistent with the estimates of other authors, which lie within the range
$9^\circ-16^\circ$ (Vall\'ee 2017b; Nikiforov and Veselova 2018; Reid et al. 2019).

Figure 5 presents the radial velocities of masers as
a function of Galactocentric distance and the power
spectrum of the radial velocities. Figure 6 presents
the residual tangential velocities of masers and their
power spectrum.

An analysis of the present-day data shows that in a wide vicinity of R0 the values of $f_R$ and $f_\theta$  are
typically 4--9 km s$^{-1}$, while the wavelength $\lambda$ lies within the range 2--3 kpc. Of special interest are the determinations of these parameters separately from the radial and tangential velocities of stars.
Having analyzed $\sim$200 Cepheids from the Hipparcos catalogue (1997), Bobylev and Bajkova (2012)
found $f_R=6.8\pm0.7$ km s$^{-1}$ and $f_\theta=3.3\pm0.5$ km s$^{-1}$, $\lambda=2.0\pm0.1$ kpc, and $\chi_\odot=-193^\circ\pm5^\circ$.

Having analyzed the spatial distribution of a large sample of classical Cepheids, Dambis et al. (2015)
estimated the pitch angle of the spiral pattern $i=-9.5^\circ\pm0.1^\circ$ and the Sun’s phase $\chi_\odot=-121^\circ\pm3^\circ$ for the four-armed spiral pattern.

\begin{figure}[t]
{\begin{center}
   \includegraphics[width=0.9\textwidth]{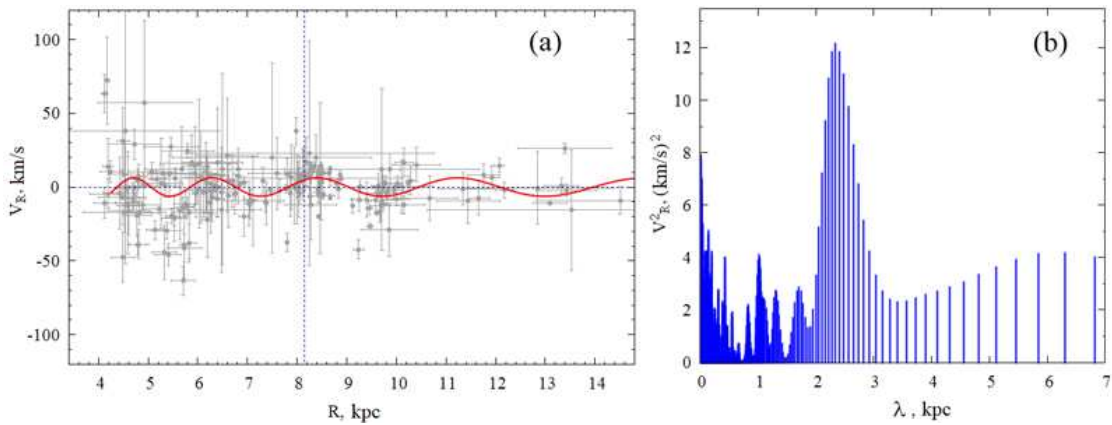}
 \caption{
(a) Maser radial velocity versus Galactocentric distance, the vertical dotted line marks the Sun’s position. (b) The power spectrum of the radial velocities.
  } \label{f-Rad}
\end{center}}
\end{figure}
\begin{figure}[t]
{\begin{center}
   \includegraphics[width=0.9\textwidth]{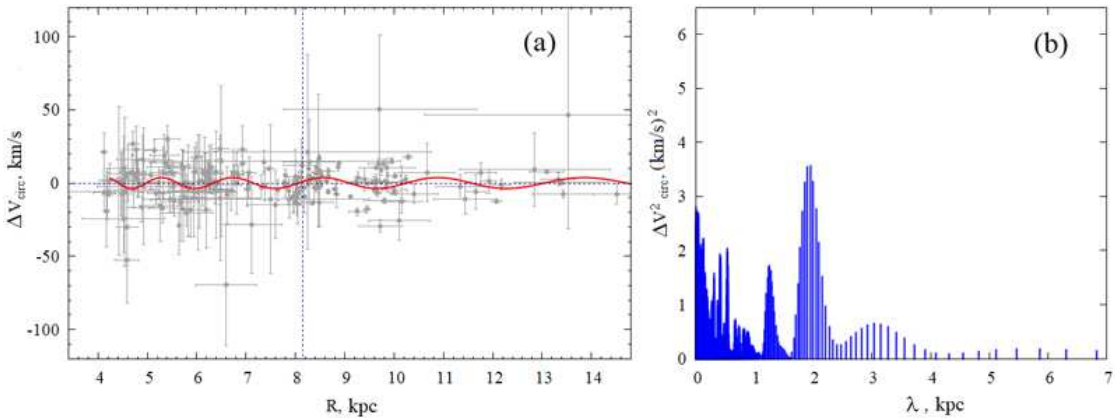}
 \caption{
(a) Maser residual tangential velocity versus Galactocentric distance, the vertical dotted line marks the Sun’s position. (b) The power spectrum of the residual tangential velocities.
  } \label{f-Tan}
\end{center}}
\end{figure}

Based on 130 maser sources with measured trigonometric parallaxes, Rastorguev et al. (2017)
found $f_R=6.9\pm1.4$ km s$^{-1}$ and $f_\theta=2.8\pm1.0$ km s$^{-1}$ and the Sun's phase $\chi_\odot=-125^\circ\pm10^\circ$. Based on $\sim$500 OB stars from the Gaia DR2 catalogue, Bobylev and Bajkova (2018) found $f_R=7.1\pm0.3$ km s$^{-1}$ and $f_\theta=6.5\pm0.4$ km s$^{-1}$, $\lambda_R=3.3\pm0.1$ kpc and $\lambda_\theta=2.3\pm0.2$ kpc, and $(\chi_\odot)_R=-135^\circ\pm5^\circ$ and
 $(\chi_\odot)_\theta=-123^\circ\pm8^\circ$. Note also the new values of $f_R=4.6\pm0.7$ km s$^{-1}$ and
$f_\theta=1.1\pm0.4$ km s$^{-1}$ obtained recently by Loktin and Popova (2019) by analyzing the present-day data on open star clusters.

Thus, the velocity perturbations $f_R$ and $f_\theta,$  the wavelengths $\lambda_R$ and $\lambda_\theta,$ and the Sun's phases in the spiral density wave $(\chi_\odot)_R$ and $(\chi_\odot)_\theta$ obtained in this paper are in good agreement with the determinations of these parameters by other authors.

 \subsection*{CONCLUSIONS}
Based on published data, we produced a sample of 256 radio sources whose trigonometric parallaxes
and proper motions were measured by VLBI. Galactic maser sources associated with massive protostars
and stars located in active star-forming regions constitute the overwhelming majority of the sample.
VLBI measurements of a number of young low-mass stars from the Gould Belt region observed in continuum
were also included in our sample.

At $R<4$ kpc the influence of the Galactic bar on the stellar space velocities is very strong. This leads to
a high circular velocity dispersion in this zone. Therefore, for our kinematic analysis we took objects in the
range of distances $R:$ 4--15 kpc. Based on this sample of 239 sources, we estimated the velocities $U_\odot,V_\odot,W_\odot$ and the parameters of the angular velocity of Galactic rotation $\Omega_0,\Omega^{(1)}_0,\ldots,\Omega^{(4)}_0$ and obtained a new estimate of $R_0=8.15_{-0.20}^{+0.04}$ kpc. For our further analysis of the residual tangential velocities $\Delta V_{circ}$ we chose the rotation curve with three derivatives $(n=3).$

Based on the series of radial, $V_R,$ and residual tangential, $\Delta V_{circ}$, stellar velocities, we found the parameters of the Galactic spiral density wave by applying a periodogram analysis. The amplitudes of
the radial and tangential velocity perturbations are $f_R=7.0\pm0.9$ km s$^{-1}$ and $f_\theta=3.8\pm1.1$ km s$^{-1}$, the perturbation wavelengths are $\lambda_R=2.3\pm0.2$~kpc and $\lambda_\theta=2.0\pm0.4$ kpc for the adopted four-armed spiral pattern $(m=4).$ The Sun's phases in the spiral density wave are $(\chi_\odot)_R=-163^\circ\pm9^\circ$ and $(\chi_\odot)_\theta=-137^\circ\pm10^\circ$.

  \bigskip{\bf REFERENCES}{\small

 1. I. Ablimit, G. Zhao, C. Flynn, and S. A. Bird, Astrophys. J. Lett. 895, 12 (2020).

 2. R. Abuter, A. Amorim, N. Baub\"ock, et al. (Gravity Collab.), Astron. Astrophys. 625, L10 (2019).

 3. V. V. Bobylev and A. T. Bajkova, Astron. Lett. 38, 638 (2012).

 4. V. V. Bobylev and A. T. Bajkova, Astron. Lett. 39, 809 (2013).

 5. V. V. Bobylev and A. T. Bajkova, Mon. Not. R. Astron. Soc. 473, 1549 (2014a).
 
 6. V. V. Bobylev and A. T. Bajkova, Mon. Not. R. Astron. Soc. 441, 142 (2014b).

 7. V. V. Bobylev and A. T. Bajkova, Astron. Lett. 40, 783 (2014c).
 
 8. V. V. Bobylev and A. T. Bajkova, Astron. Lett. 44, 675 (2018).

 9. A. G. A. Brown, A. Vallenari, T. Prusti, et al. (Gaia Collab.), Astron. Astrophys.
616, 1 (2018).

 10. T. Camarillo, M.Varun, M. Tyler, and R. Bharat, Publ. Astron. Soc. Pacif. 130, 4101 (2018).

 11. A. K. Dambis, L. N. Berdnikov, Yu. N. Efremov, et al., Astron. Lett. 41, 489 (2015).

 12. T.Do, A. Hees,A. Ghez, G.D. Martinez, D. S. Chu, S. Jia, S. Sakai, J. R. Lu, et al., Science (Washington,
DC, U. S.) 365, 664 (2019).

 13. P. A. B. Galli, L. Loinard, G. N. Ortiz-Le\'on, M. Kounkel, S. A. Dzib, A. J. Mioduszewski,
L. F. Rodriguez, L. Hartmann, et al., Astrophys. J. 859, 33 (2018).

 14. R. de Grijs and G. Bono, Astrophys. J. Suppl. Ser. 232, 22 (2017).

 15. The Hipparcos and Tycho Catalogues, ESA SP--1200 (1997).

 16. T. Hirota, T. Nagayama, M. Honma, Y. Adachi, R. A. Burns, J. O. Chibueze, Y. K. Choi,
K. Hachisuka, et al. (VERA Collab.), arXiv:2002.03089 (2020).

 17. M. Honma, T. Nagayama, and N. Sakai, Publ. Astron. Soc. Jpn. 67, 70 (2015).

 18. M. Honma, T. Nagayama, T. Hirota, N. Sakai, T. Oyama, A. Yamauchi, I. Toshiaki, T. Handa, et al.,
in Maser Astrometry and Galactic Structure Study with VLBI, Proc. IAU Symp. 336, 162 (2018).

 19. M. Kounkel, L. Hartmann, L. Loinard, G. N. Ortiz-Le\'on, A. J. Mioduszewski, L. F. Rodriguez,
S. A. Dzib, R. M. Torres, et al., Astrophys. J. 834,142 (2017).

 20. C. C. Lin and F. H. Shu, Astrophys. J. 140, 646 (1964).

 21. A. V. Loktin and M. E. Popova, Astrophys. Bull. 74, 270 (2019).

 22. I. I. Nikiforov and A. V. Veselova, Astron. Lett. 44, 81 (2018).
 
 23. G. N. Ortiz-Le\'on, L. Loinard, S. A. Dzib, P. A. B. Galli, M. Kounkel, A. J. Mioduszewski,
L. F. Rodriguez, R. M. Torres, et al., Astrophys. J. 865, 73 (2018).

 24. G. N. Ortiz-Le\'on, K. M. Menten, T. Kaminski, A. Brunthaler, M. J. Reid, and R. Tylenda, Astron.
Astrophys. 638, 17 (2020).

 25. T. Prusti, J.H. J. de Bruijne,A.G. A. Brown, A. Vallenari, C. Babusiaux, C. A. L. Bailer-Jones, U. Bastian, M. Biermann, et al. (Gaia Collab.), Astron. Astrophys. 595, A1 (2016).

 26. A. S. Rastorguev, M. V. Zabolotskikh, A. K. Dambis,
N. D. Utkin, V. V. Bobylev, and A. T. Baikova, Astrophys.
Bull. 72, 122 (2017).

 27. M. J. Reid and M. Honma, Ann. Rev. Astron. Astrophys. 52, 339 (2014a).

 28. M. J. Reid, K. M. Menten, A. Brunthaler, X.W. Zheng, T.M. Dame, Y. Xu, Y. Wu, B. Zhang, et
al., Astrophys. J. 783, 130 (2014b).

 29. M. J.Reid, K.M. Menten, X.W. Zheng, and A. Brunthaler, Astrophys. J. 823, 77 (2016).

 30. M. J. Reid, N. Dame, K. M. Menten, A. Brunthaler,
X.W. Zheng, Y. Xu, J. Li, N. Sakai, et al., Astrophys.
J. 885, 131 (2019).

 31. N. Sakai, H. Nakanishi, M. Matsuo, N. Koide, D. Tezuka, T. Kurayama, K.M. Shibata, Y. Ueno, and
M. Honma, Publ. Astron. Soc. Jpn. 67, 69 (2015).

 32. N. Sakai, T. Nagayama, H. Nakanishi, N. Koide,
T. Kurayama, N. Izumi, T. Hirota, T. Yoshida, et al.,
arXiv:1910.08146 (2019).

 33. R. Sch\"onrich, J. Binney, and W. Dehnen, Mon. Not.
R. Astron. Soc. 403, 1829 (2010).

 34. J. P. Vall\'ee, Astrophys. Space Sci. 362, 79 (2017a).

 35. J. P. Vall\'ee, New Astron. Rev. 79, 49 (2017b).

 36. Y. Xu, J. J. Li, M. J. Reid, K.M. Menten, X.W. Zheng,
A. Brunthaler, L. Moscadelli, T. M. Dame, and
B. Zhang, Astrophys. J. 769, 15 (2013). 
  }
  \end{document}